\documentclass[%
 reprint,
 amsmath,amssymb,
 aps,
pra,
]{revtex4-2}

\usepackage{graphicx}
\usepackage{dcolumn}
\usepackage{bm}
\usepackage{color}
\usepackage{hyperref}

\usepackage{soul}
\usepackage{xcolor}
\sethlcolor{yellow}
\usepackage{tcolorbox}
\soulregister\cite7

\begin{document}
\preprint{APS/123-QED}
\title{Spin Correlations in Recirculating Multipass Alkali Cells for Advancing Quantum Magnetometry}

\author{Qian Ling Kee\textsuperscript{1,2}}
\thanks{Current affiliation: Cavendish Laboratory, University of Cambridge, JJ Thomson Ave Cambridge CB3 0HE, United Kingdom} \author{Lingyi Zhao\textsuperscript{3}}
\author{Ruvi Lecamwasam\textsuperscript{1,2}}
\author{Biveen Shajilal\textsuperscript{1,2}}
\author{Xinan Liang\textsuperscript{2}}
\author{Joel K Jose\textsuperscript{1,2}}
\author{Yao Chen\textsuperscript{4}}
\author{Ping Koy Lam\textsuperscript{1,2,5,6}}
\email{Corresponding author: pingkoy@a-star.edu.sg}
\author{Tao Wang\textsuperscript{1,2}}
\email{Corresponding author: tao\_wang@a-star.edu.sg}

\affiliation{\textsuperscript{1}Quantum Innovation Centre (Q.InC), Agency for Science Technology and Research (A*STAR), 2 Fusionopolis Way, Innovis \#08-03, Singapore 138634, Republic of Singapore}
\affiliation{\textsuperscript{2}Institute of Materials Research and Engineering (IMRE), Agency for Science Technology and Research (A*STAR), 2 Fusionopolis Way, Innovis \#08-03, Singapore 138634, Republic of Singapore}
\affiliation{\textsuperscript{3}Xi'an Jiaotong University, No.28 Xianning West Road, Xi'an, Shaanxi 710049, P.R. China}
\affiliation{\textsuperscript{4}School of Instrument Science and Technology, Xi’an Jiaotong University, No.28 Xianning West Road, Xi’an 710049, P.R. China}
\affiliation{\textsuperscript{5}Centre for Quantum Technologies, National University of Singapore, 3 Science Drive 2, Singapore, 117543, Republic of Singapore}
\affiliation{\textsuperscript{6}Centre for Quantum Computation and Communication Technology, Department of Quantum Science and Technology, Australian National University, ACT 2601, Australia}

\vspace{10pt}

\begin{abstract}
Multipass cells enable long optical path lengths in compact volumes and are central to quantum technologies such as atomic magnetometers and optical quantum memories. In optical magnetometry, multipass geometries enhance sensitivity by increasing optical depth, reducing photon shot noise, and enabling quantum non-demolition detection. However, in conventional cylindrical multipass cells, Lissajous beam trajectories lead to repeated revisiting and incomplete mirror coverage, limiting effective volume utilization. Here we present a recirculating multipass alkali cell that overcomes these limitations by increasing the active-to-cell volume ratio and minimizing beam spot overlap. We develop an analytical ABCD-matrix model to predict beam trajectories, spot sizes, and astigmatism, validated by Zemax simulations. We further introduce a general analytical framework for spin correlation noise that incorporates astigmatism and spatial intensity distributions. By deriving the spin-noise time-correlation function and spectrum, we show how beam intensity profiles influence spin diffusion noise. Our results demonstrate improved beam coverage, reduced spot overlap, and enhanced spin correlation, particularly for concave mirrors with long focal lengths, while showing that avoiding tightly-focused regions significantly suppresses spin diffusion noise. These findings establish recirculating multipass cells as a practical, high-performance platform for precision atomic sensing and other multipass-cavity–based quantum devices.

\vspace{1pc}

\end{abstract}
%

%
%
\maketitle
%
%


\section{Introduction}
Multipass cells have a wide range of applications that enhance their performance, including the post-compression of ultrashort laser pulses \cite{viotti2022multi,jargot2018self}, molecular beam absorption spectroscopy \cite{kaur1990multipass}, and increased measurement sensitivity in micro-fluidic devices \cite{argueta2024increasing}. Moreover, they play a crucial role in quantum technologies and are widely used in optical magnetometers and optical quantum memories to improve performance \cite{arnold2023free, lvovsky2009optical, Sheng2013}. Herriott multipass cells, which offer a compact design while providing long optical path lengths, can extend the storage time of optical quantum memories by several orders of magnitude \cite{arnold2024all,das2011very}. In atomic magnetometers, a single-pass alkali cell allows probe light to traverse the alkali vapor only once, limiting its interaction with the medium. In contrast, a multipass cell enables the light to pass through the vapor multiple times, increasing optical absorption and polarization rotation without requiring higher atomic densities or larger physical cells. This design enhances the optical rotation signal, reduces noise, and improves the signal-to-noise ratio, ultimately leading to higher magnetometer sensitivity. Understanding the spatial distribution of multipass beams is generally essential — for instance, when investigating laser beam interactions with atoms. To address this, we propose a comprehensive analytical model and Python code to fully reproduce the multipass beam spatial distribution. This is useful for optimizing the use of multipass cells in various applications particularly in compact setups, where achieving sufficient sensitivity would otherwise be challenging.

For instance, scalar atomic magnetometers, which measure the Larmor precession frequency, can operate effectively in finite magnetic fields, offering metrological advantages such as higher fractional resolution and inherent self-calibration \cite{smullin2009low}. These magnetometers are employed in a wide range of applications, including biomedical sensing in ambient environments \cite{Limes2020}, magnetic anomaly detection \cite{zhao2021brief, lu2023recent}, magnetic resonance imaging (MRI) \cite{savukov2013anatomical}, space exploration \cite{bennett2021precision}, and magnetic navigation \cite{bulatowicz2012compact}. However, their performance in finite fields is fundamentally limited by spin-exchange relaxation, which shortens spin relaxation time. Although this effect can be partially suppressed using light narrowing techniques \cite{appelt1999light, scholtes2011light, han2017light}, complete elimination is not possible. Multipass cells offer a promising solution by enabling quantum non-demolition (QND) measurements in scalar magnetometers under a pulsed pump-probe regime, enhancing sensitivity without increasing the measurement volume. A key factor enabling QND measurements is high optical depth, which multipass cells significantly improve. For example, sub-femtotesla sensitivity has been demonstrated using two cylindrical multipass alkali cells in scalar magnetometers \cite{Sheng2013,lucivero2021femtotesla}. Moreover, a fast rotating field (FRF) vector magnetometer, combining a fast rotating magnetic field with a multipass scalar magnetometer, was developed. It achieved a fractional resolution of 0.7 parts per billion and angular sensitivities of 6~$\mathrm{nrad/\sqrt{Hz}}$ for two polar angles in Earth’s magnetic field, significantly enhancing the accuracy of magnetic source localization \cite{wang2025pulsed, wang2026pulsed}.

Under ideal technical conditions, the sensitivity of atomic magnetometers is fundamentally limited by quantum noise sources, including spin-projection noise ($\delta B_{spn}$), photon shot noise ($\delta B_{psn}$), and atomic spin diffusion ($\delta B_{asd}$) \cite{ledbetter2008spin,seltzer2008developments,mouloudakis2023interspecies,dellis2014spin,tang2020spin,lucivero2017correlation,shaham2020quantum}. Atomic spin diffusion arises from the thermal motion of atoms, leading to fluctuations in their spatial distribution and time-dependent variations in the detected spin signal. This effect is further influenced by atomic diffusion in and out of the detection region, which can be analyzed through the spin noise correlation function.  The quantum noise contributions scale as $\delta B_{spn} \sim \frac{1}{\gamma} \sqrt{\frac{1}{n_v V T_2}}$ and $\delta B_{psn} \sim \frac{1}{\gamma} \frac{1}{T_2 \sqrt{n_v V \Gamma_{pr} O\!D_0}}$, where $\gamma$ is the gyromagnetic ratio, $T_2$ is the transverse relaxation time, $n_v$ is the vapor density, $V$ is the active volume, $\Gamma_{pr}$ is the probe rate, and $O\!D_0 = n_v \sigma l$ is the optical depth. As a key parameter in QND measurements \cite{shah2010high,takahashi1999quantum,Auzinsh2004}, $O\!D_0$ can be significantly enhanced in multipass cells through increased effective interaction length, thereby improving magnetometer sensitivity.

Among various multipass Herriott cell designs, cylindrical cells are most commonly used for atomic magnetometry. However, in conventional cylindrical multipass cells, the transverse beam cross-section follows a Lissajous pattern \cite{Silver2005,li2011optical,liu2022femtotesla} that does not fully cover the mirror surfaces and repeatedly revisits the same regions, reducing the effective active volume and limiting miniaturization of high-sensitivity atomic magnetometers. Moreover, unavoidable tightly focused regions accelerate the decay of spin correlations due to atomic diffusion and increase spin noise \cite{Sheng2013}.

To address these limitations, we propose a recirculating multipass alkali cell based on two planar mirrors and one concave mirror. This design improves spatial beam coverage and reduces spot overlap, thereby enhancing the active-to-cell volume ratio and enabling more uniform interaction between the probe light and atomic ensemble. We present an analytical model of the recirculating multipass cell, consistent with simulations using commercial ray-tracing software, and provide a publicly accessible Python program to facilitate the design of multipass cells for applications in optical quantum memories and magnetometry.

In conventional cylindrical cells, the non-uniform Lissajous beam distribution leads to an effective filling factor on the order of $2/\pi \sim 0.64$ or lower, due to repeated sampling and unfilled regions. In contrast, the recirculating geometry approaches a more uniform coverage of the available volume, resulting in a larger effective active volume $V$. As a consequence, a greater number of atoms contribute independent fluctuations, and the corresponding quantum noise scales favorably as $\delta B_{spn} \propto (n_v V)^{-1/2}$ and $\delta B_{psn} \propto (n_v V O\!D_0)^{-1/2}$. This increase in active volume reduces both spin-projection and photon shot noise through statistical averaging. 

Furthermore, an analytical model for spin noise in multipass alkali cells has been lacking. For example, in Sheng’s work, the analysis relies primarily on numerical simulations rather than a closed-form treatment \cite{Sheng2013}. While an analytical model has been developed for single-pass geometries \cite{lucivero2017correlation}, it does not account for multipass configurations or the astigmatism of Gaussian beams. Here, we develop a general analytical framework for spin noise in multipass alkali cells that explicitly incorporates beam astigmatism and spatial intensity distributions, and is applicable to a broad class of geometries. Using this model, we analyze a recirculating multipass cell—a configuration not previously studied—and show that spin noise decreases with increasing focal length, offering advantages for high-precision quantum sensing. We further demonstrate that recirculating designs can outperform conventional cylindrical cells. In addition, while prior studies have qualitatively suggested that tightly focused regions may increase spin noise, our results provide the first quantitative confirmation of this effect.

\section{Theory of Spin Noise in Atomic Vapor Cells}

The sensitivity of atomic magnetometers is fundamentally limited by spin projection noise, which imposes a quantum limit as dictated by Heisenberg’s uncertainty principle. Additionally, spin noise due to diffusion introduces an extra layer of classical noise, further degrading sensitivity. This makes a thorough understanding of spin noise essential. Given the rich correlations present in atomic vapors, correlation functions are crucial for capturing the temporal and spatial dynamics of atomic spin noise \cite{mouloudakis2023interspecies}. By the Wiener-Khinchin theorem, the spin noise power spectral density $S(f)$ of the optical rotation signal can be expressed as the Fourier transform of its time autocorrelation. This relationship simplifies to
\begin{equation}
S(f)  = \left\langle\phi(t)^2\right\rangle \int_{0}^{\infty} 2C(|\tau|) e^{-i2 \pi f \tau} \, d \tau,
\label{eq:fourier}
\end{equation}
where $S(f)$ is expressed in terms of the spin noise time-correlation function $C(\tau)$, which is given by the normalized time autocorrelation of the Faraday rotation $\phi$ of the probe beam
\begin{equation} 
C(\tau)=\frac{\left\langle \phi(t)\phi(t+\tau) \right\rangle}{\left\langle \phi(t)^2 \right\rangle}.
\label{eq:c}
\end{equation}
This correlation function forms the basis for understanding how spin noise is influenced by spin projection-induced Faraday rotation, which we discuss in the following section.

\subsection{Spin Noise Time-Correlation Function}
We start by deriving the spin noise time-correlation function resulting from atomic diffusion to quantify the decay of atomic spin correlations over time. Given that the probe beam propagates along the $\hat{z}$ direction and thereby measures atomic spins projected onto the $z$-axis, we focus on the spin expectation value in $z$, denoted as $\langle s_z \rangle$. The correlation of $\langle s_z \rangle$ at positions $\mathbf{r}_1$ and $\mathbf{r}_2$ at any given time $t$ is given by
\begin{equation} 
\left\langle s_z\left(\mathbf{r}_1, t\right) s_z\left(\mathbf{r}_2, t\right)\right\rangle_F=\left\langle s_z^2\right\rangle_F \frac{\delta\left(\mathbf{r}_1-\mathbf{r}_2\right)}{n_v},
\label{eq:covequaltime}
\end{equation}
where $F$ is the total atomic spin and $\delta\left(\mathbf{r}_1-\mathbf{r}_2\right)$ is the Dirac delta function which represents spatial coincidence at $\mathbf{r}_1=\mathbf{r}_2$, so $\langle s_z \rangle$ for randomly polarized atoms are correlated only at the same position. This correlation equals the variance of the spin expectation value $\left\langle s_z^2\right\rangle_F$ given by $\left\langle s_z^2\right\rangle_F=\frac{2 F+1}{2(2 I+1)} \frac{1}{(2 I+1)^2} \frac{F(F+1)}{3}.$ After introducing a time lag $\tau$, the evolution of atomic spins in vapor cells with buffer gas is described by 
\begin{align}
(s_z + i s_x)(\mathbf{r}_2, \tau)
&= e^{-i \omega_L \tau - \tau / T_2}
\int G(\mathbf{r}_1 - \mathbf{r}_2, \tau)
\nonumber \\
&\quad \times (s_z + i s_x)(\mathbf{r}_1, 0)\,
d^3 \mathbf{r}_1 .
\label{eq:spin}
\end{align}
where $s_z$ and $s_x$ denote $z$- and $x$- components of atomic spin polarisation respectively. Here, the spin dynamics depend on the spin precession at Larmor frequency $\omega_L$, the transverse spin relaxation time $T_2$, and the Green's function describing the spatial distribution of diffusing spins over the time lag, which is given by $G(\mathbf{r}_1 - \mathbf{r}_2, \tau) = (4D\tau)^{-3/2}e^{-\frac{|\mathbf{r}_1 - \mathbf{r}_2|^2}{4D\tau}}$, where $D$ is the diffusion constant.

From Eq.~\ref{eq:spin}, we can derive the correlation of spin expectation values at $\mathbf{r}_1$ and at $\mathbf{r}_2$ given a time lag $\tau$, which simplifies to
\begin{equation}
\begin{aligned}
& \left\langle s_z(\mathbf{r}_1, t)\,
s_z(\mathbf{r}_2, t+\tau)\right\rangle_F
= \cos(\omega_L \tau)\, e^{- \tau / T_2}
\\
&\int G(\mathbf{r}_3 - \mathbf{r}_2, \tau)
\times
\left\langle s_z(\mathbf{r}_1, t)\,
s_z(\mathbf{r}_3, t)\right\rangle_F\,
d^3 \mathbf{r}_3 .
\label{eq:spintimelagfull}
\end{aligned}
\end{equation}
where $\mathbf{r}_3$ is introduced as a dummy variable for the integration to distinguish it from $\mathbf{r}_1$ and $\mathbf{r}_2$. As the spin expectation values $\langle s_z(\mathbf{r}_1, t) \rangle$ and $\langle  s_z(\mathbf{r}_3, t) \rangle$ are the variables that need to be averaged, the constants $\cos(\omega_L \tau), e^{-\tau / T_2}, $ and $G(\mathbf{r}_3 - \mathbf{r}_2, \tau)$ are independent of these spin variables and can therefore be factored out of the averaging operation $\left\langle \hat{O} \right\rangle_F$. Substituting Eq.~\ref{eq:covequaltime} into Eq.~\ref{eq:spintimelagfull} gives 
\begin{equation}
\begin{aligned}
& \left\langle s_z(\mathbf{r}_1, t)\,
s_z(\mathbf{r}_2, t+\tau)\right\rangle_F
= \cos(\omega_L \tau)\, e^{- \tau / T_2} \\
&\int G(\mathbf{r}_3 - \mathbf{r}_2, \tau)
 \times
\left\langle s_z^2 \right\rangle_F
\frac{\delta(\mathbf{r}_1-\mathbf{r}_3)}{n_v}\,
d^3 \mathbf{r}_3,
\end{aligned}
\end{equation}
which simplifies to
\begin{equation}
\left\langle s_z\left(\mathbf{r}_1, t\right) s_z\left(\mathbf{r}_2, t+\tau\right)\right\rangle_F=\cos(\omega_L \tau)e^{- \tau / T_2} G(\mathbf{r}_1 - \mathbf{r}_2, \tau) \frac{\left\langle s_z^2\right\rangle_F}{n_v}.
\label{eq:spintimelag}
\end{equation} 
since $\delta\left(\mathbf{r}_1-\mathbf{r}_3\right)$ is only non-zero at $\mathbf{r}_1=\mathbf{r}_3$. This expression represents a generalized form of Eq.~\ref{eq:covequaltime} as it includes the effects given a finite time lag $\tau$. Intuitively, this reflects how atoms initially at $\mathbf{r}_1$ diffuse to $\mathbf{r}_2$ over time $\tau$, causing the spin at $\mathbf{r}_1$ to become correlated with the spin at $\mathbf{r}_2$ at the time lag $\tau$. 

\subsection{Optical Rotation Correlation}
Since spin projection fluctuations of diffusion-driven atomic spins along the probe beam affect the polarization rotation of an off-resonant, linearly polarized probe beam, we now explore how spin projection fluctuations manifests as optical rotation. Specifically, when there is a spin projection along the probe beam, it induces birefringence, leading to a measurable rotation of the probe beam’s polarization \cite{wu1986optical}. The paramagnetic Faraday rotation of the probe beam is given by \cite{happer1967off}
\begin{equation}
\phi (t) = \sum_{F=I \pm 1 / 2} \frac{c r_e f_{\mathrm{osc}} n_v~\mathrm{Im}(\mathcal{L}(\nu-\nu_F))}{\int I(\mathbf{r}) d x d y}  \int I(\mathbf{r}) \langle s_z \rangle (\mathbf{r},t)  \,d^3 \mathbf{r},
\label{eq:opticalrotation}
\end{equation} 
where $c$ is speed of light in vacuum, $r_e$ is the radius of electron, $f_{osc}$ is the oscillator strength and $I(\mathbf{r})$ is intensity distribution of the probe beam. The magnetometer response is given by the imaginary part of a Lorentzian profile $\mathrm{Im}(\mathcal{L}(\nu-\nu_F)) = \frac{(\nu-\nu_F)}{\left(\nu-\nu_F\right)^2+\Gamma^2}$, which is an approximation of the Voigt profile where sufficient buffer gas is present and pressure broadening dominates the lineshape. Here, $\nu$ denotes the probe beam frequency, $\nu_F$ represents the resonance frequencies of the D1 and D2 transitions and $\Gamma$ is the pressure broadening linewidth.

From Eq.~\ref{eq:opticalrotation}, we can obtain the time autocorrelation of rotation which is given by

\begin{align}
&\langle \phi(t)\phi(t+\tau)\rangle
= \sum_{F=I \pm 1/2}
\left(
\frac{c r_e f_{\mathrm{osc}} n_v\,
\mathrm{Im}\!\left[\mathcal{L}(\nu-\nu_F)\right]}
{\int I(\mathbf{r})\, dx\, dy}
\right)^2
\nonumber \\
&\quad \times
\int I(\mathbf{r}_1)\, I(\mathbf{r}_2)\,
\left\langle
s_z(\mathbf{r}_1, t)\,
s_z(\mathbf{r}_2, t+\tau)
\right\rangle_F\, d^3 \mathbf{r}_1\, d^3 \mathbf{r}_2 .
\label{eq:autocovariance}
\end{align}

Similarly, from Eq.~\ref{eq:opticalrotation}, the variance of the optical rotation can be derived as
\begin{equation}
\begin{aligned}
 \left\langle \phi(t)^2 \right\rangle & =\sum_{F=I \pm \frac{1}{2}} \left( \frac{c r_e f_{\mathrm{osc}} n_v~\mathrm{Im}(\mathcal{L}(\nu-\nu_F))}{\int I(\mathbf{r}) \, dx \, dy} \right)^2 \\
 & \int I(\mathbf{r})^2 \frac{\left\langle s_z^2\right\rangle_F}{n_v} \, d^3 \mathbf{r}.   
\end{aligned}
\label{eq:rotation variance}
\end{equation}
Substituting Eqs.~\ref{eq:autocovariance} and \ref{eq:rotation variance} into Eq.~\ref{eq:c}, we obtain the time correlation function:
\begin{equation}
C(\tau)=\frac{\int I(\mathbf{r_1}) I(\mathbf{r_2}) G(\mathbf{r_1}-\mathbf{r_2},\tau)\cos(\omega_L\tau) e^{-\tau/T_2} d^3\mathbf{r}_1 d^3\mathbf{r}_2}{\int I(\mathbf{r})^2 d^3\mathbf{r}}.
\end{equation} 
with the diffusion component $C_d(\tau)$ is given by
\begin{equation}
C_d(\tau) = \frac{\int I(\mathbf{r}_1) I(\mathbf{r}_2) G(\mathbf{r}_1 - \mathbf{r}_2, \tau) \,d^3\mathbf{r}_1 \,d^3\mathbf{r}_2}{\int I(\mathbf{r})^2 \,d^3\mathbf{r}}.
\label{eq:Cd}
\end{equation}
$C_d(\tau)$ can be understood as the likelihood that atoms initially in the probe beam remain within the beam after time $\tau$, even if they diffuse out and later return. It depends solely on the intensity distribution of the probe beam and the Green’s function. Therefore, to determine $C_d(\tau)$ for multipass cells, we must first characterize the intensity distribution of the probe beam within such cells. Specifically, we focus on the recirculating multipass cell \cite{Robert2007}, which features a high number of passes to reduce photon shot noise and a large active volume that enhances both photon shot noise and spin projection noise performance. 

\section{Atomic Vapour Cell based on Recirculating Multipass Cavity}
\begin{figure*}[htbp]
\centering
	\includegraphics[width=18cm]{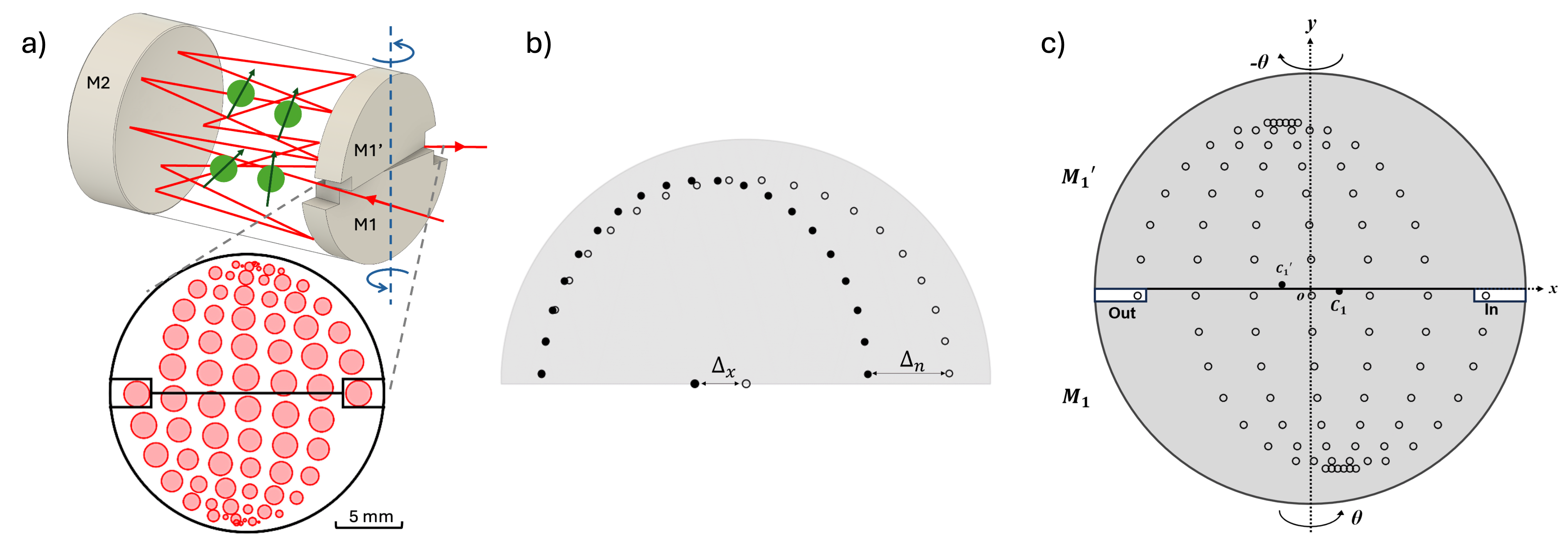}
	\caption{a) Illustration of the recirculating multipass alkali cell (top) with an incident beam entering from one side of $M_1$ and exiting from the opposite side. The internal beam paths (red lines) are simplified for clarity and do not represent the actual number of passes or beam paths. Arrowed spheres indicate atomic spins. The mirrors $M_1$ and $M_1^{\prime}$ are rotated about the y-axis (blue dotted line) to create a recirculating beam reflection pattern (bottom) observed on $M_1$ and $M_1^{\prime}$. The simulation parameters are $\theta_x = 0.04^{\circ}$, $\theta_x' = -0.04^{\circ}$, $x_0=8.11$~mm, $y_0=0$~mm, $x_0'=-0.26^{\circ}$, $y_0'=2.21^{\circ}$, $d=29.8$~mm, and $f=1$~m, which gives 78 reflections. b) The beam spots positions on $M1'$, with hollow circles marking their positions before the mirror's rotation and filled circles indicating their positions after the rotation. The tilting of the mirror results in an offset to the optical center $\Delta_x$ and an offset to the last reflected spot after a 180$^\circ$ turn about the optical center $\Delta_n$. c) Beam spot distributions on $M_1$ and $M_1^{\prime}$ rotated in opposite directions such that the beam spots circulate around $C_1^{\prime}$ on $M_1^{\prime}$ and around $C_1$ on $M_1$, forming a recirculating pattern.}
	\label{fig:schematic}
\end{figure*}

To determine the spin noise, we first determine the optical path within the recirculating cell. The recirculating multipass cell, based on the Herriott cell design, can be understood as multiple Herriott cells within a single, compact cavity. As shown in Fig.~\ref{fig:schematic}a, it comprises three mirrors:  $M_1$  and  $M_1^{\prime}$, each with focal length  $f_1$, and  $M_2$,  with a focal length  $f_2$. Mirrors $M_1$ and $M_1^{\prime}$ are positioned at distance $d$ from $M_2$, and they form the lower and upper halves, respectively. They are tilted in opposite directions by small angles $\theta_x$ (for $M_1$) and $\theta_x'$ (for $M_1^{\prime}$) about the central axis of the mirror pair, aligned along the $y$-axis, forming a recirculating optical pattern with substantially more passes than the typical Herriott cell. To examine the distribution of beam spots in the recirculating cell, we calculate their positions on the mirrors $M_1$  and  $M_1^{\prime}$.

\subsection{Beam Spot Distribution of Recirculating Multipass Cell}

We analyze the distribution of beam spots within the recirculating multipass cell using ray transfer matrix analysis. As the ray travels between the mirrors repeatedly, the transfer matrix of the ray within the multipass cell can be simply described using the matrix for a single round trip:
\begin{equation}
\left[\begin{array}{cc} A & B \\ C & D \end{array}\right] =
\left[\begin{array}{cc} 1 & 0 \\ -1/f_1 & 1 \end{array}\right]
\left[\begin{array}{cc} 1 & d \\ 0 & 1 \end{array}\right]
\left[\begin{array}{cc} 1 & 0 \\ -1/f_2 & 1 \end{array}\right]
\left[\begin{array}{cc} 1 & d \\ 0 & 1 \end{array}\right].
\label{eq:ABCD}
\end{equation}
such that the ray enters from $M1$, travels a distance $d$ to $M2$, reflects off $M2$, then returns to $M1$ and reflects off it. The y-position $y_n$ and $y$ slope $y_n'$ of the $n^\mathrm{th}$ reflection off M1 can be described using the transfer matrix as
\begin{equation}
\left[\begin{array}{c} y_n \\ y_n' \end{array}\right] =
\left[\begin{array}{cc} A & B \\ C & D \end{array}\right]^n
\left[\begin{array}{c} y_0 \\ y_0' \end{array}\right],
\end{equation}
where $y_0$ and $y_0'$ are the initial $y$-position and $y$ slope. $y_n$ can be written in the form
\begin{equation}\label{eq:y0}
y_n = y_0 \cos(n\theta) + \sqrt{2df_2-d^2} y_0' \sin(n\theta)= Y \sin(n\theta + \beta),
\end{equation}
where $\theta=\arccos\left(\frac{A+D}{2}\right)=\arccos(1-d/f_2)$ is the angle $\theta$ between the beam spots on the mirrors. $\theta$ depends solely on the focal length $f_2$ of the mirror and the distance $d$ between the mirrors and does not change with the position and angle of the incident light.
The $x$-position $x_n$ and $x$ slope $x_n'$  can be described in a similar way when $M_1$  and  $M_1^{\prime}$ are not tilted, and $x_n$ can be written in the form
\begin{equation}
x_n = x_0 \cos(n\theta) + \sqrt{2df_2-d^2} x_0' \sin(n\theta)= X \sin(n\theta + \alpha).\label{eq:x0}\end{equation}
where $x_0$ and $x_0'$ are the initial $x$-position and $x$ slope. The quantities $X$ and $Y$ are the maximum amplitude of the beam spot in the $x$- and $y$- direction respectively. Without considering the tilt in $M1$ and $M1'$, Eqs.~\ref{eq:y0} and \ref{eq:x0} describe the beam spots on the mirrors of a typical Herriott cell, which traces an elliptical path.

However, $x_n$ and $x_n'$ differ in the recirculating multipass cell due to the tilt of $M_1$  and  $M_1^{\prime}$, which introduces an offset of the optical center. This offset is key to enabling the beam spots to recirculate to achieve a higher number of passes compared to Herriott cells. Given a small rotation of $\theta_x$ of a mirror, $x$ and $x'$ changes such that
\begin{equation}
\left[\begin{array}{cc} x_s \\ x_s' \end{array}\right] \to 
\left[\begin{array}{c} x_s \\ x_s' \end{array}\right] + 
\left[\begin{array}{c} 0 \\ \theta_x \end{array}\right],
\end{equation}
where $s$ is a specific identifier indicating the incident beam on the rotated mirror. Since the angular components of the parameters are relative to the normal of the mirror surface, and in a rotated mirror system, the normal also rotates by the same angle $\theta_x$, it is necessary to include an additional term to ensure that the ray transfer matrix parameters are referenced back to the original coordinate system. Given that $M_1$  and  $M_1^{\prime}$ are plane mirrors, $ f_1 \to \infty $), we can say that
\begin{equation}
\label{ABCDtransform}
\left[\begin{array}{cc} 1 & 0 \\ -\frac{1}{f_1} & 1 \end{array}\right] 
\biggl(\left[\begin{array}{c} x_s \\ x_s' \end{array}\right] + \left[\begin{array}{c} 0 \\ \theta_x \end{array}\right]\biggr) 
+ \left[\begin{array}{c} 0 \\ \theta_x \end{array}\right] = 
\left[\begin{array}{c} x_s \\ x_s' \end{array}\right] + \left[\begin{array}{c} 0 \\ 2\theta_x \end{array}\right].
\end{equation}
Thus, a single round trip is given by
\begin{equation}
\label{eq:xsABCD}
\left[\begin{array}{c} x_{s+1} \\ x_{s+1}' \end{array}\right] =
\left[\begin{array}{cc} A & B \\ C & D \end{array}\right]
\left(\left[\begin{array}{c} x_s \\ x_s' \end{array}\right] + 
\left[\begin{array}{c} 0 \\ 2\theta_x \end{array}\right]\right),
\end{equation}
which describes a recursive relationship where a perturbation term is added for each round trip due to reflection from a tilted mirror. Consequently, the parameters after $n$  reflections from the rotated mirror can be expressed as
\begin{equation}
\left[\begin{array}{c} x_{s+n} \\ x_{s+n}' \end{array}\right] =
\left[\begin{array}{cc} A & B \\ C & D \end{array}\right]^n
\left[\begin{array}{c} x_s \\ x_s' \end{array}\right] + 
\sum_{i=1}^n \left[\begin{array}{cc} A & B \\ C & D \end{array}\right]^i
\left[\begin{array}{c} 0 \\ 2\theta_x \end{array}\right],
\label{eq:xparameters}
\end{equation}
where the first term represents the ray characteristics of an non-rotated mirror, while the second term sums up the perturbations arising from the rotated mirror. This equation relies on the specific case that simplifies Eq.~\ref{ABCDtransform} -- mirror $M1$ is flat, with a focal length that is effectively infinite.

The y-offset, $\Delta_n$, where the subscript $n$ indicates reflection number, corresponds to the second term in Eq.~\ref{eq:xparameters}. We can derive the following expression:
\begin{equation}
\Delta_n=B\sum_{i=1}^n U_{i-1}\times2\theta_x,
\label{eq:y_offset}
\end{equation}
where
\begin{equation}
U_N = \frac{\sin{(N+1)\theta}}{\sin{\theta}}, 
\end{equation}
and
\begin{equation}
\theta = \arccos\left(\frac{A+D}{2}\right).
\label{eq:theta}
\end{equation}
By substituting specific parameters, Eq.~\ref{eq:y_offset} can be simplified to
\begin{equation}\label{eq:OFFSET}\Delta_n=2\theta_x\times\sqrt{2df-d^2}\sum_{i=1}^n\sin i \theta,\end{equation}
which is the offset for each beam spot reflected off the mirror. This offset accumulates with increasing reflection number, as shown in Fig.~\ref{fig:schematic}b.

Notably, the offset of the last reflected spot (after a 180$^\circ$ turn about the optical center) $\Delta_n$ is approximately twice the offset of the optical center $\Delta_x$ such that $\Delta_x\approx2\Delta_n$. Substituting this into Eq.~\ref{eq:OFFSET}, we express the offset in the optical center as
\begin{equation}\label{eq:centerOFFSET}\Delta_x\approx\theta_x\times\sqrt{2df_2-d^2}\sum_{i=1}^n\sin i \theta. \end{equation}
To calculate the $x$-position of the beam spots on the recirculating multipass cell, we can simply add the offset to the initial position to account for the tilt of the mirrors. If the beam spot rotates anti-clockwise upon laser incidence into the cavity, the position offset is will be first influenced by mirror $M_1$. At this point, the position of the beam spot can be expressed as shown in Eq.~\ref{eq:xn}, where the summation term $ m_0 $ is the difference between the parameter $ n $ for the desired position calculation and the parameter $ n_0 $ for the first reflection on $M_1$.
\begin{equation}
x_n = x_0 \cos(n\theta)+2\theta_x\times\sqrt{2df-d^2}\sum_{i=1}^{m_0}\sin i \theta.
\label{eq:xn}
\end{equation}

After a 180$^\circ$ anti-clockwise circulation about the optical center, the beam now enters $M_1^{\prime}$. As there is a rotation in $M_1^{\prime}$ relative to $M_1$ at angle of $ \theta_1' - \theta_1 $, now the summation term  is instead $ m_1 $ which is the difference between the parameter $ n $ for the desired calculation and the parameter $ n_1 $ for the first reflection on the second affected mirror $M_1^{\prime}$. It subsequently undergoes further circulations, making this a recirculating multipass cell. As the beam spots recirculate , they alternate between $M_1$ and $M_1^{\prime}$. We define the integer $k$ to describe the circulation number of a beam spot where k is
\begin{equation}
k=\lfloor \frac{n\theta}{\pi} \rfloor.
\end{equation}
We derive $k$ by taking the floor function of the reflection angle parameter divided by $\pi$. The floor function ensures that $k$ takes integer values. This computation allows $k$ to quantify the number of recirculation cycles encountered by the light within the optical system. Additionally, the parity of $k$—whether it is odd or even—indicates whether the beam is on $M_1$ or $M_1^{\prime}$. Considering that the beam spots alternate between $M_1$ and $M_1^{\prime}$, we modify Eq.~\ref{eq:xn} to arrive at the general $x$-position of the $n$-th beam spot given by
\begin{equation}\label{eq:xn_final}
x_n = x_0 \cos(n\theta) +2\sqrt{2df-d^2}\times\sum_{j=0}^{k}\sum_{i=1}^{m_j}\theta_j\sin i \theta,
\end{equation}
where each change of the reflection mirror adds another summation term to the position description, and each term’s summation index $ m_j $ is the difference between the parameter $ n $ for the desired calculation and the parameter $ n_j $ for the first reflection on the corresponding mirror. Here, $\theta_j$ represents the relative rotation angle of the mirror, taking values of $\theta_x$, $-2\theta_x$, $2\theta_x$, $-2\theta_x$, $2\theta_x$, and so on. Therefore, we obtain recirculating beam spot positions that indicates spatially distributed beams through the alkali cell, which increases active volume of measurement.

As shown in Fig.~\ref{fig:schematic}c, mirrors $M_1$ and $M_1^{\prime}$ rotate in opposite directions by the same angle, denoted as $\Delta_x=-\Delta_x'=\Delta$. This causes the optical center offset of $M_1$ to be in the positive x-direction, and $M_1^{\prime}$ in the negative direction. The incident light enters at the position $\left(x_0, 0\right)$ with an angle of $\left(0, \varphi_{y0}\right)$, where $y_0'$ is set to a negative angle, enabling the light spots to rotate clockwise. The schematic diagram below specifically illustrates the configuration of $M_1$ and $M_1^{\prime}$  within the cavity design. This configuration extends the optical path length of the probe beam through the alkali cell while spatially distributing the beam across the cell. Furthermore, since the optical path length of the recirculating cell affects the spin noise dynamics of the atomic vapor, we also aim to determine the total number of allowed reflections in this type of cell, as it is proportional to the path length.

\subsection{Total Number of Reflections}
Next, we determine the total number of possible reflections on the mirror. In our configuration, the initial beam spot on $M_1$ originates from $(x_0, 0)$ and first passes through the $M_1$ section, where the beam center is at $(\Delta, 0)$. This trajectory intersects the x-axis again at $(2\Delta - x_0, 0)$. Subsequently, the light passes through the $M_1^{\prime}$ section, where the beam center is at $(- \Delta, 0)$. Thus, after one complete cycle, the light returns to $(x_0 - 4\Delta, 0)$. Following this pattern, after $N$ cycles, the position can be represented as $(x_0 - 4N\Delta, 0)$. Given the symmetrical nature of this design, the criterion for the exit condition is that the position must be less than $-x_0$. Therefore, the total number of recirculation required to meet this condition can be calculated based on the inequality:
\begin{equation}x_0 - 4N\Delta<-x_0.\end{equation}
Since the number of recirculation, $N$, must be an integer, it can be calculated using the ceiling function. With $N$ recirculation, the total angular displacement of the light spots amounts to $2\pi N$. Considering Eq.~\ref{eq:theta}, where the angle between two consecutive reflection points is $\theta$, the total number of reflections $n_{R}$ on $M_1$ can be expressed as follows:
\begin{equation}\label{eq:total}
n_{R}=\frac{2\pi N}{\theta}=\frac{2\pi}{\theta}\left\lceil\frac{x_{0}}{2\Delta}\right\rceil. 
\end{equation}

\begin{figure*}[htbp]
  \centering
  \includegraphics[width=15cm]{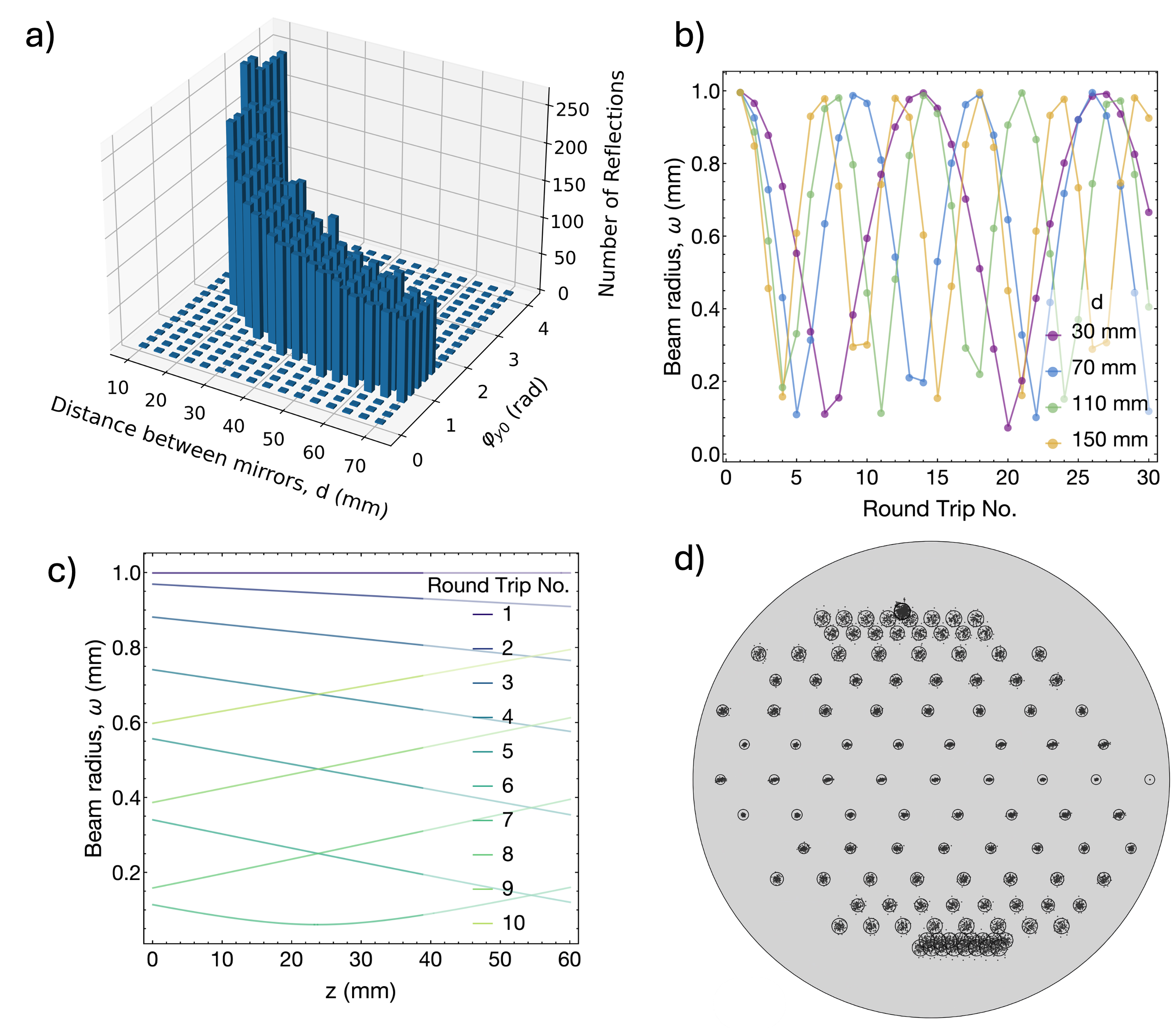} 
  \caption{a) Calculated number of reflections in a recirculating cell as a function of the distance between mirrors and input angle in the y-direction at $\lambda = 780~\mathrm{nm}$, $\omega_0 = 1~\mathrm{mm}$, $\theta_x = 0.02^\circ$, $\varphi_{x0} = 0^\circ$, and $f_2 = 1~\mathrm{m}$. The results show that the recirculating design is insensitive to mirror misalignment. b) Trace of beam radius along $z$ for each round trip at d = 30 mm. Beam traverse twice the mirror separation for a round trip. c) Oscillating evolution of beam radius with number of round trips for various mirror separations $d$.  d) Analytical beam positions and waist (hollow circles) and Zemax simulated light rays (scatter points). The multipass cell parameters are: $f_2=1000~\mathrm{mm}$, M1 is a plane mirror, $d=86.46~\mathrm{mm}$, the rotation angles $\theta_x$ of mirrors $M_1$ and $M_1^{\prime}$ are 0.02$^\circ$ and -0.02$^\circ$, respectively. The incident beam has position $ x_0 = 11 $~mm and $ y_0 = 0 $~mm, and angles $ \varphi_{x0} = 0^\circ$, and $ \varphi_{y0} = 1.2^\circ $. Calculated beam radius from Eq.~\ref{eq:qn} and Eq.~\ref{eq:beamwidth} with $f_2$ = 1000~$\mathrm{mm}$ and $\omega_0=1~\mathrm{mm}$.}
  \label{fig:cellgeometry}
\end{figure*}

Furthermore, the total number of reflections depends on both the initial input angle in the $y$ direction, $y_0'$, and the distance between the mirrors, as shown in Fig.~\ref{fig:cellgeometry}a. For each mirror separation, there exists a range of allowable $y_0'$ values over which the number of reflections remains unchanged, reflecting the sensitivity to variations in $y_0'$. In particular, $y_0'$ can vary by large angle without affecting the total number of reflections, demonstrating the robustness of the recirculating design against mirror misalignment. In general, the required $y_0'$ increases as the mirror separation decreases. Shorter mirror separations yield a higher number of reflections, which, while beneficial for increasing the interaction length, also indicate increased sensitivity to small variations in mirror separation. As the mirror distance increases, the $y_0'$ required to achieve a high number of reflections stabilizes, saturating at approximately $1.25~\mathrm{mm}$.

\subsection{Beam Width}
In addition to the total number of allowed reflections, the beam width also influences the spin noise of the atoms in the multipass cell. To account for this effect, we derive the beam width throughout the cell, allowing us to calculate the beam’s intensity distribution. This, in turn, enables us to determine $C_d$ in Eq.~\ref{eq:Cd}. Assuming that the probe beam propagates as a Gaussian beam, the complex beam parameter q at $M_1$ and $M_1^{\prime}$ changes with reflection number $n$ such that
\begin{equation}
q_n=\frac{Aq_0+B}{Cq_0+D},
\label{eq:qn}
\end{equation}
where the ABCD matrix is given by Eq.~\ref{eq:ABCD}. The initial beam parameter is $q_0=z+\frac{i\pi\omega_0^2}{\lambda}$ with position $z$ defined relative to beam waist $\omega_0$ and wavelength $\lambda$.
In general, the complex beam parameter at position $z$ is given by
$\frac{1}{q(z)} = \frac{1}{R(z)} - \frac{i \lambda}{\pi \zeta w(z)^2}$ where $R$ is the radius of curvature of the beam's wavefront, $\zeta$ is the refractive index of the medium, and $w$ is the beam radius, which is defined as the distance from the beam center at which the intensity falls to $1/e^2$ ($\sim$13.5\%) of its maximum value on the axis. It is given by $w(z) = w_0 \sqrt{1 + \left( \frac{z}{z_R} \right)^2}$ where $z_R$ is the Rayleigh range, defined as $z_R = \frac{\pi w_0^2}{\lambda}$. In the case that that the first reflection is positioned at the beam waist of the initial beam, the initial beam has $ z = 0 $ set at $ \omega_0 = 1 $ mm. The beam radius of the $ n $-th successive reflection on M1 or $M_1^{\prime}$ is given by
\begin{equation}
\omega_n = \sqrt{-\frac{\lambda}{\pi \mathrm{Im}\left(\frac{1}{q_n}\right)}}.
\label{eq:beamwidth}
\end{equation}


Fig.~\ref{fig:cellgeometry}b shows the change in beam radius with varying mirror separations for the first 30 round trips, calculated from Eq.~\ref{eq:beamwidth} for a plane mirrors M1 and $M_1^{\prime}$, and a concave mirror M2 with $f_2$ = 1~$\mathrm{m}$. The beam radius varies periodically with successive reflections, with a larger mirror separations resulting in shorter periods of oscillation. The beam is also focusing with increasing round trips due to concave mirror $M2$ and then expands after a certain number of round trips as shown in Fig.~\ref{fig:cellgeometry}c.

\subsection{Validating Analytical Model against Ray-tracing Simulation Results}
To validate the analytical calculations, we compare the predicted beam positions and widths with those obtained from commercial ray-tracing simulations (Zemax in our case). We simulate a Gaussian beam by using 100 light rays as represented by the black scatter points on the mirror surface in Fig.~\ref{fig:cellgeometry}d. The black hollow circles indicate the beam positions and width calculated analytically using the same parameters as the simulation. These parameters also give a $\theta = 0.4189^\circ$. This implies that in a single recirculation, there are $2\pi/\theta \approx 15$ reflections, with 7 reflections occurring on a single rotated mirror. The offset of the optical center $\Delta$, calculated using Eq.~\ref{eq:OFFSET}, is $0.6754~\mathrm{mm}$. Substituting this into Eq.~\ref{eq:total}, we find the total number of reflections $n_{R}$ to be 120, which is consistent with the simulation results.

Furthermore, to validate the analytical beam positions, we calculate the average position of the rays to determine the spot location. This approach was chosen because it reduces random errors associated with the position of individual rays, thereby providing a more accurate representation of the actual spot location. We find that the average error distance for the 120 reflection positions is 0.0122~$\mathrm{mm}$. 

We also observe that the beam width evolution follows the expected periodic variations in Fig.~\ref{fig:cellgeometry}b. We further validate our calculation by computing the proportion of the light rays within the beam width at each beam spot. For a Gaussian beam, the spots within the beam width should contain 87\% of the total number of spots. Indeed, the proportion of light spots within the beam width converges towards 87\% with increasing number of rays simulated, demonstrating the validity of our analytical model. Our model can also provide the distribution of the beam waist size as the probe beam propagates through the cell.


\section{Spin Noise in Multipass Cells}
Finally, we propose a model to derive the spin-noise diffusion correlation function for a general multipass cell by solving Eq.~\ref{eq:Cd} using the intensity distribution of the multipass beams. It is essential to account for beam ellipticity caused by astigmatism, particularly from reflections off cylindrical mirrors, as they introduce asymmetry in the beam cross-section. The intensity distribution of an elliptical Gaussian beam is given by
\begin{equation}
    I(\xi,\eta,z)=\left| E_\xi(\xi,z) E_\eta(\eta,z)\right|,
    \label{eq:intensityastigmatic}
\end{equation}
where $E$ is the relative field strength of an elliptical Gaussian beam, and $\xi$ and $\eta$ are the lengths along the principal axes. For a circular Gaussian beam, $E_\xi=E_\eta$, hence its intensity distribution can be simplified to
\begin{equation}
    I=\left| E(r,z) \right|^2.
    \label{eq:I(r)}
\end{equation}
where $r$ is the radius of the beam cross-section. In general, $E(r)$ is expressed as
\begin{equation}
    E(r,z) = \frac{1}{\sqrt{q(z)}} \exp \left( -i k \frac{r^2}{2 q(z)} \right),
    \label{eq:u(r)}
\end{equation}
where $k$ is the wavenumber of the light. By propagating the beam in $z$ after each reflection on $M1$, we obtain the complex beam parameter for the $n^{th}$ pass  which is given by $q(z)=q_n +z$ such that $q_n$ is the beam parameter on $M1$ and $z$ is the probe beam's travel within that round trip. $q_n$ is given by
\begin{equation}
    q_n=\mathbf{R}^n q_0
\end{equation}
for the round trip matrix $\mathbf{R}$ of any multipass cell. For ease of derivation, we simplify Eq.~\ref{eq:u(r)} by taking $\frac{1}{q(z)}=a-ib$ where $a$ and $b$ are real numbers, giving $E(r,z)=\sqrt{a^2+b^2} e^{-kr^2b/2}e^{-ikr^2a/2}e^{-i \arctan(b/a)}$. Solving Eq.~\ref{eq:I(r)}, the intensity distribution can expressed in a more concise form:
\begin{equation}
    I (\mathbf{r})= (a^2+b^2)e^{-kr^2b}.
    \label{eq:intensity}
\end{equation}

We will now solve Eq.~\ref{eq:Cd} by evaluating the numerator and denominator separately. Using Eq.~\ref{eq:intensity}, the numerator of Eq.~\ref{eq:Cd}, $\int I(\mathbf{r}_1) I(\mathbf{r}_2) G(\mathbf{r}_1 - \mathbf{r}_2, \tau) \, d^3\mathbf{r}_1 \, d^3\mathbf{r}_2$, for the $n^{th}$ round trip can be expressed as
\begin{equation}
\begin{aligned}
= & \int (a_1^2 + b_1^2)e^{-k(x_1^2 + y_1^2)b_1} (a_2^2 + b_2^2) e^{-k(x_2^2 + y_2^2)b_2} \\
& \frac{1}{(4D\tau)^{3/2}} e^{-\frac{(x_1 - x_2)^2 + (y_1 - y_2)^2 + (z_1 - z_2)^2}{4D\tau}}  \, d^3\mathbf{r}_1\, d^3\mathbf{r}_2. 
\end{aligned}
\end{equation}
Assuming that the cell size is much larger than beam width, we can subsequently integrate with respect to $x$ and $y$ from $-\infty$ to $\infty$ to obtain
\begin{equation}
 = \int_{-d}^{d} \int_{-d}^{d} \frac{\sqrt{\pi}(a_1^2 + b_1^2)(a_2^2 + b_2^2)}{2k\sqrt{D\tau} (b_1+b_2+4b_1b_2Dk\tau)}  e^{-\frac{(z_1 - z_2)^2}{4D\tau}} \, dz_1 \, dz_2.
    \label{eq:numerator}
\end{equation}
To further simplify the evaluation, we substitute $u=z_1-z_2$ and $v=z_1+z_2$ into Eq.~\ref{eq:numerator} to obtain
\begin{equation}
\begin{aligned}
&= \int_{-2d}^{0}\int_{-2d-u}^{2d+u} 
\frac{\sqrt{\pi}(a_1^2 + b_1^2)(a_2^2 + b_2^2)}{4k\sqrt{D\tau} (b_1+b_2+4b_1b_2Dk\tau)} e^{-u^2/4D\tau} \, dv \, du \\ 
& + \int_{0}^{2d}\int_{-2d+u}^{2d-u} 
\frac{\sqrt{\pi}(a_1^2 + b_1^2)(a_2^2 + b_2^2)}{4k\sqrt{D\tau} (b_1+b_2+4b_1b_2Dk\tau)} e^{-u^2/4D\tau} \, dv \, du.
\label{eq:subuv}
\end{aligned}
\end{equation}

Since the spin noise at $z_1$ is only weakly correlated with that at $z_2$ when the distance $|z_1 - z_2|$ is much greater than the cell length, we can assume  $u \ll d$. Taking $u=0$ everywhere except at the Gaussian function $e^{-u^2/4D\tau}$, which is an even function, the numerator becomes
\begin{equation}
    = \int_{-2d}^{2d} \frac{\sqrt{\pi}(a_1^2 + b_1^2)(a_2^2 + b_2^2)}{2k\sqrt{D\tau} (b_1+b_2+4b_1b_2Dk\tau)} \,dv \int_{0}^{2d} e^{-u^2/4D\tau} \,du.
\end{equation}
Using the definition of the error function, we find that $\int_{0}^{2d} e^{-u^2/4D\tau} \,du=\sqrt{\pi D\tau } \times \mathrm{erf}(\frac{d}{\sqrt{D\tau}})$, which can also be written in terms of the complementary error function: $\mathrm{erf}(\frac{d}{\sqrt{D\tau}})=1-\mathrm{erfc}(\frac{d}{\sqrt{D\tau}})$. For large $d$, we can apply the asymptotic expansion giving $\mathrm{erfc}(\frac{d}{\sqrt{D\tau}}) \approx \frac{\sqrt{D\tau}}{d\sqrt{\pi}}e^{-d^2/D\tau}$. As $d$ is large and $D$ is small, this expression tends to 0 so we arrive at the simplified numerator:
\begin{equation}
    = \sqrt{\pi D \tau}\int_{-2d}^{2d} \frac{\sqrt{\pi}(a_1^2 + b_1^2)(a_2^2 + b_2^2)}{2k\sqrt{D\tau} (b_1+b_2+4b_1b_2Dk\tau)} \,dv.
    \label{eq:numeratorfinal}
\end{equation}
The denominator of Eq.~\ref{eq:Cd}, $\int I(\mathbf{r})^2 \,d^3\mathbf{r}$, for the $n^{th}$ round trip can be expressed as
\begin{equation}
    =\int_{-d}^{d} (a^2+b^2)^2 \frac{\pi}{2bk} \, dz.
    \label{eq:denominator}
\end{equation}
Finally, we can evaluate the normalised spin noise diffusion correlation Eq.~\ref{eq:Cd} for multipass cells by taking the sum of in Eq.~\ref{eq:numeratorfinal} over all passes divided by the sum of Eq.~\ref{eq:denominator} over all passes. 

We can solve Eq.~\ref{eq:Cd} in a similar manner for astigmatic Gaussian beams, but using Eq.~\ref{eq:intensityastigmatic} instead. Since the focal points of the two principal axes do not coincide in astigmatic beams, we propagate two distinct $q$-parameters, $q_\xi$ and $q_\eta$. We then compute $E_\xi$ and $E_\eta$ using Eq.~\ref{eq:u(r)}, which allows us to obtain Eq.~\ref{eq:intensityastigmatic}. The numerator of Eq.~\ref{eq:Cd}, $\int I(\mathbf{r}_1) I(\mathbf{r}_2) G(\mathbf{r}_1 - \mathbf{r}_2, \tau) \, d^3\mathbf{r}_1 \, d^3\mathbf{r}_2$, can then be derived to give 
\begin{widetext}
\begin{equation}
\begin{aligned}
 =&\int_{-d}^{d} \int_{-d}^{d} \frac{\pi}{k} 
\frac{1}{\sqrt{b_\xi b_\eta \left( 2 a_\xi^2 + z_1^2 + z_2^2 + 2 a_\xi (z_1 + z_2) 
+ 2 b_\xi \left( b_\xi + 2 D k \tau \right) \right)}} \\ 
&  \times \frac{1}{\sqrt{\left( 2 a_\eta^2 + z_1^2 + z_2^2 + 2 a_\eta (z_1 + z_2) 
+ 2 b_\eta \left( b_\eta + 2 D k \tau \right) \right)}} \, dz_1 \, dz_2,
\label{eq:numeratorastig}
\end{aligned}
\end{equation}
\end{widetext}
and the denominator, $\int I(\mathbf{r})^2 \,d^3\mathbf{r}$, is given by
\begin{equation}
 =\int_{-d}^{d} \frac{\pi}{2k \sqrt{b_\xi b_\eta}} \cdot \frac{1}{ \sqrt{ \left[ (a_\xi + z)^2 + b_\xi^2 \right] \left[ (a_\eta + z)^2 + b_\eta^2 \right] } } \, dz.
\label{eq:denominatorastig}
\end{equation}
Similarly, we can evaluate the normalised spin noise diffusion correlation Eq.~\ref{eq:Cd} for astigmatic Gaussian beams in multipass cells by taking the sum of in Eq.~\ref{eq:numeratorastig} over all round trips divided by the sum of Eq.~\ref{eq:denominatorastig} over all round trips. With $\mathbf{R}$ defined for any multipass cell, we can now derive the spin-noise diffusion correlation function in multipass cells. In the Supplementary Information (Fig.~S1), we validate our spin noise model by demonstrating good agreement with the cylindrical cell calculation reported by Sheng et al. \cite{Sheng2013} and the single pass calculation by Vito et al. \cite{lucivero2017correlation}.

\subsection{Spin Noise in Recirculating Multipass Alkali Cells}
\begin{figure*}[htbp]
  \centering
   \includegraphics[width=18cm]{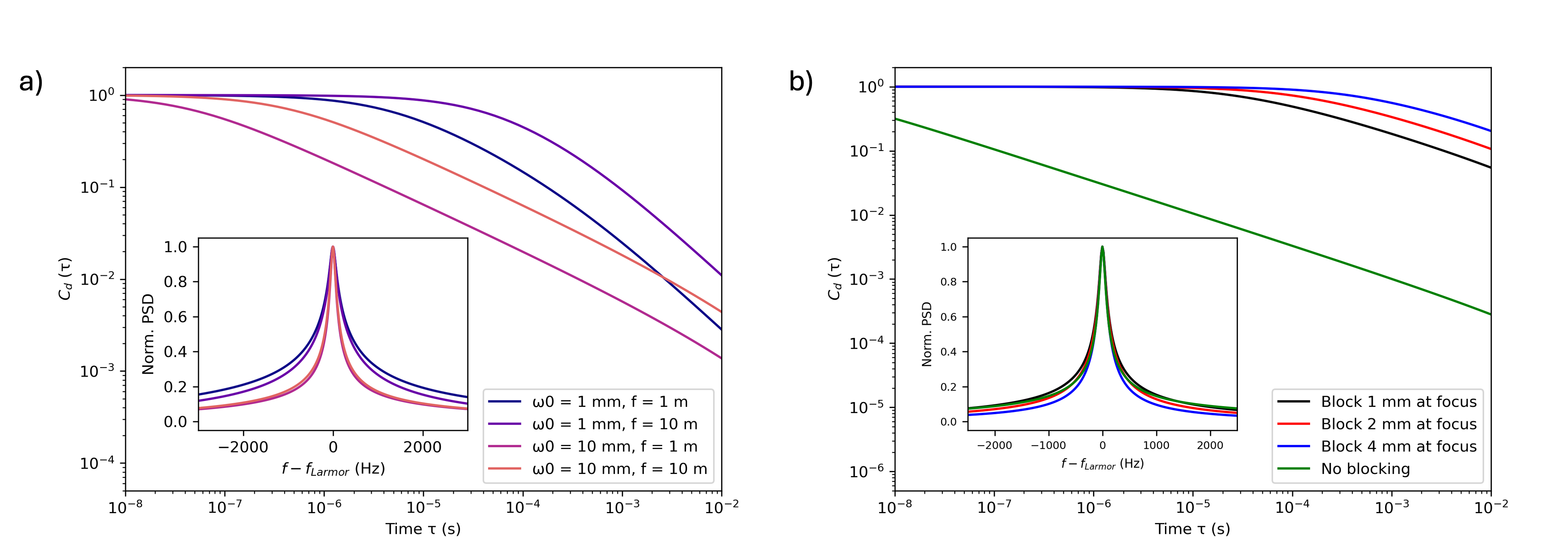}
  \caption{a) Normalized spin noise diffusion correlation $C_d$ of recirculating cell calculated from Eq.~\ref{eq:numeratorfinal} and Eq.~\ref{eq:denominator} and normalized power spectral density calculated from \ref{eq:fourier}. The calculation is performed for $\omega_0 = 1~\mathrm{mm}$ and $\omega_0 = 10 ~\mathrm{mm}$, and focal lengths $f=1~\mathrm{mm}$ and $f=10~\mathrm{mm}$. The cell parameters are d~=~30~mm and n~=~50. b) $C_d$ and normalised PSD (inset) for a single pass with $f~=$~1.3~m and $d=$~45~mm and $\omega_0$~=~1~mm. By inserting a barrier to prevent atoms from entering the tightly focused region, the spin correlation function exhibits a slower decay, and the spin diffusion noise is reduced (inset).}
  \label{fig:recir_spinnoise}
\end{figure*}

To study the spin noise in the recirculating multipass cell in Eq.~\ref{eq:Cd}, we take round trip matrix $\mathbf{R}$ from Eq.~\ref{eq:ABCD}. For a beam reflecting off a concave mirror, the focal length in the tangential and sagittal direction is $f_{tan}=f\cos\phi$ and $f_{sag}=f/\cos\phi$ respectively where $\phi$ is the angle between the beam and the normal of incidence \cite{Paschottafocal_length}. For small $\phi$, $\Delta f = f_{tan} - f_{sag} \approx f\phi^2/2\approx 0$. Therefore, the beams in the recirculating multipass cell can be considered approximately stigmatic. In Fig.~\ref{fig:recir_spinnoise}a, we show the normalized spin noise diffusion correlation $C_d$ calculated from Eq.~\ref{eq:numeratorfinal} and Eq.~\ref{eq:denominator} which are applicable to stigmatic beams. The spin noise power spectral density is calculated Eq.~\ref{eq:fourier}, with the spectra normalised to their peak values. In these calculations, we take these the total number of reflections $n_R=78$, cell length $l=30~\mathrm{mm}$, $\gamma=780~\mathrm{nm}$, cell temperature $T=120^\circ\mathrm{C}$ and buffer gas pressure $70~\mathrm{Torr}$. In particular, for alkali vapor cells containing Rb atoms and $\mathrm{N}_2$ buffer gas, we can use the diffusion constant $D_0 = 0.159~\mathrm{cm}^2/\mathrm{s}$ at temperature $T_0 = 60~^\circ \mathrm{C}$ and pressure $p_0 = 760~\mathrm{Torr}$ as measured in \cite{ishikawa2000diffusion}. The diffusion constant  $D(p_{\mathrm{Rb-N_2}})$ is given by $D(p_{\mathrm{Rb-N_2}}) = D_0 \frac{p_0}{p_{\mathrm{Rb-N_2}}} \left( \frac{T}{T_0} \right)^{3/2}$ where $p_{\mathrm{Rb-N_2}}$ is the pressure of the $Rb-\mathrm{N}_2$ mixture in the alkali cell. If different alkali metal vapors, such as Na, Cs, or K, or buffer gases, such as He or Ne, are used, the diffusion constant in the calculations will need to be adjusted accordingly. The faster decay in $C_d$ for the $\omega_0=1~\mathrm{mm}$, $f=1~\mathrm{m}$ case compared to the $\omega_0=1~\mathrm{mm}$, $f=10~\mathrm{m}$ case corresponds to a larger PSD linewidth in the former case. Interestingly, a smaller initial beam waist results in a slower decay of $C_d$. This is because a larger initial beam waist leads to a higher beam divergence, resulting tighter focused region(s). 

Our analysis also reveals that regions of high optical intensity—and corresponding tight focusing—contributes significantly to spin diffusion noise. These regions, despite their strong weightage in the correlation function (as described by Eq.~\ref{eq:Cd}), limits spin coherence as their tight spatial confinement results in the quick transit of atoms through them. To mitigate this effect, we simulated the inclusion of a barrier that restricts atomic motion in these high-intensity zones for a single pass (Fig.~\ref{fig:recir_spinnoise}b). This modification led to a notable improvement in spin diffusion noise performance compared to the significant decay in $C_d$ when no blocking is introduced. Specifically, blocking even a 1~mm region around the focus significantly increases $C_d$, while further increases become progressively smaller with increasing blocking length. This indicates that the tightly focused region dominates the rapid decay of spin correlations. A small region of tight focus therefore disproportionately degrades spin-noise performance. This occurs because atoms diffuse through the high-intensity gradient near the focus on short timescales, leading to rapid loss of temporal correlation in the collective spin signal. Although tighter focusing also reduces the effective interaction volume, the dominant effect is the accelerated decay of the diffusion correlation function. Consequently, removing the focal region improves the effective correlation time and reduces spin noise. This strategy is particularly effective for systems using concave mirrors with shorter focal lengths, where the beam waist is tighter and diffusion-induced decorrelation is more pronounced.

\subsection{Comparison with Conventional Cylindrical Multipass Alkali Cells}
\begin{figure*}[htbp]
	\centering
	\includegraphics[width=15cm]{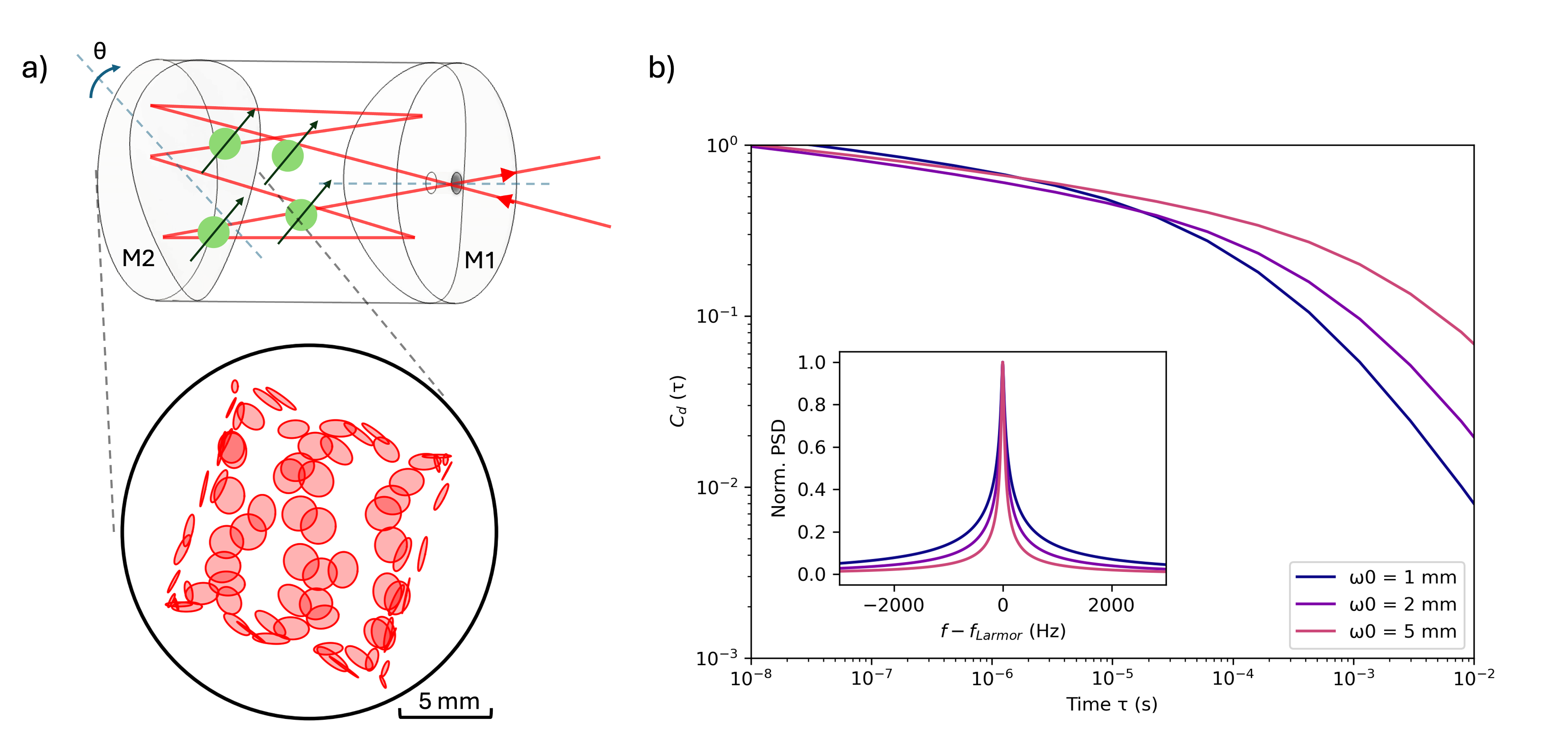}
	\caption{a) Illustration of the cylindrical multipass alkali cell (top) with an incident beam entering a hole in the center of M1 and exiting from the same hole. The internal beam paths are simplified for clarity and do not represent the actual number of passes or beam paths. Arrowed spheres indicate atomic spins. The Lissajous pattern observed on M2 (bottom) shows unavoidable large unfilled regions and tightly focused regions. The simulation parameters are twist angle $\theta=50^{\circ}$, $d=30$~mm, 78 round trips, $w_{\xi0}=w_{\eta0}=1$~mm, and $f_1=f_2=50$~mm. b) $C_d$ and normalised PSD (inset) in cylindrical cell for initial beam widths $w_{\xi0}=w_{\eta0}$ of 1~mm, 2~mm and 5~mm. The simulation parameters are twist angle $\theta=50^{\circ}$, $d=30$~mm, 21 round trips,  and $f_1=f_2=50$~mm.}
	\label{fig:cylindrical_spinnoise}
\end{figure*}

Cylindrical multipass cells have been used in scalar and vector atomic magnetometers based on fast rotating fields, demonstrating unprecedented accuracies \cite{heilman2024large,li2022kilohertz,Sheng2013,wang2025pulsed}. The basic structure of the cylindrical cell and the Lissajous beam spot distribution are shown in Fig.~\ref{fig:cylindrical_spinnoise}a. Furthermore, we calculate the spin noise in the cylindrical multipass cell, using $\mathbf{R}$ from \cite{Silver2005}. Since the beams in the cylindrical cells are astigmatic, we use Eqs.~\ref{eq:numeratorastig} and \ref{eq:denominatorastig} to calculate $C_d$. We find that the beam distribution given by the cylindrical cell gives a faster decay of $C_d$ for smaller initial beam widths as shown in Fig.~\ref{fig:cylindrical_spinnoise}b, in contrast with the case for the recirculating multipass cell in Fig.~\ref{fig:recir_spinnoise}a. 

\begin{figure*}[htbp]
  \centering
  \includegraphics[width=10cm]{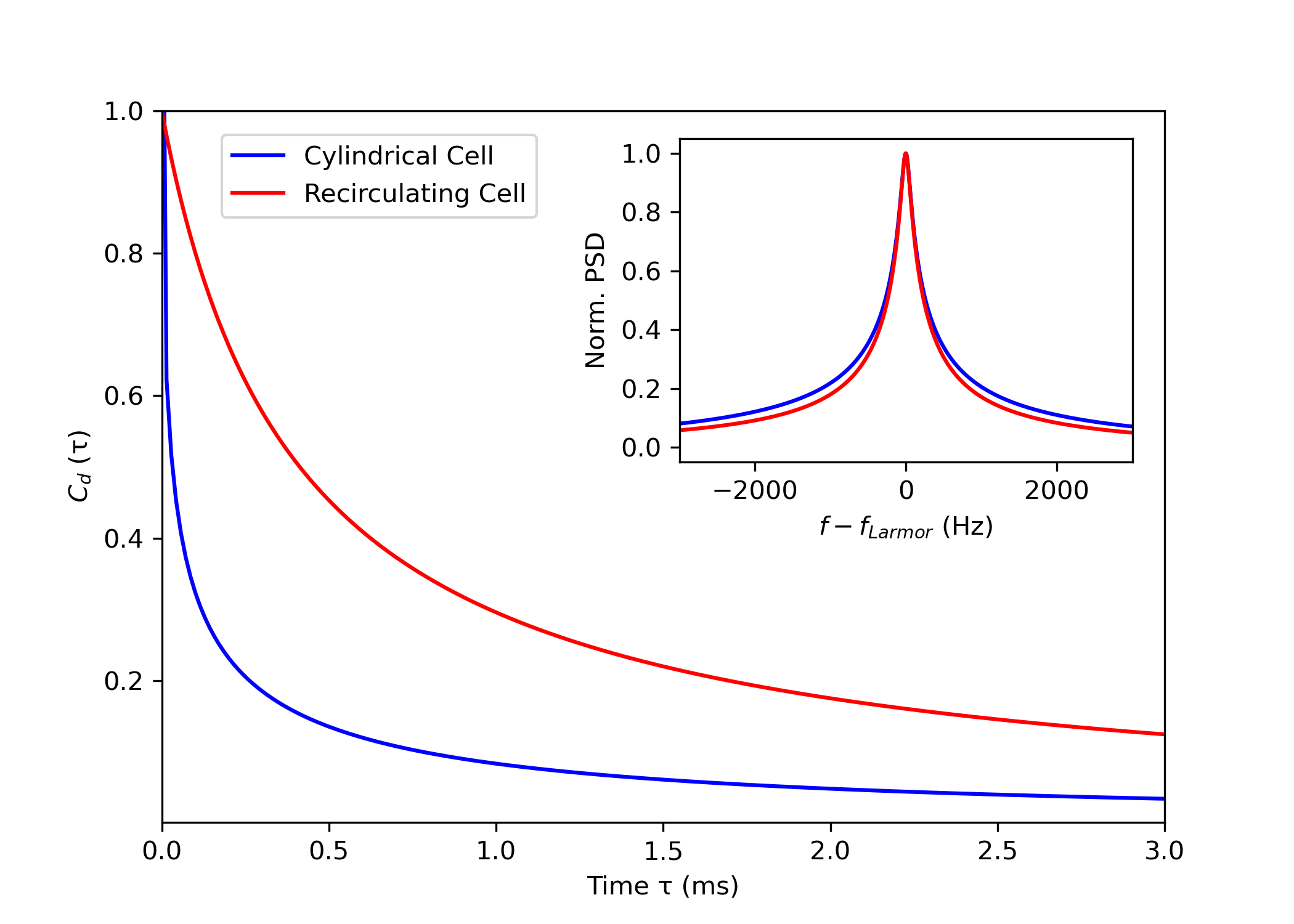}
  \caption{$C_d$ and normalised PSD (inset) of cylindrical vs recirculating multi-pass cells, both configured with 42 optical passes, $\omega_0=0.95$~mm, and $d=30$~mm. The cylindrical cell uses cylindrical mirrors with $f = 50$~mm and a twist angle of $48^\circ$, while the recirculating cell uses spherical mirrors with $f = 5$~m.}
  \label{fig:corr_compareCd}
\end{figure*}

When comparing the two types of multipass cells, the results for the cylindrical configuration are reproduced from \cite{Sheng2013}, while the recirculating multipass cell was simulated under identical conditions: 42 passes, $\omega_0=0.95$~mm, and $d=30$~mm. As shown in Fig.~\ref{fig:corr_compareCd}, the recirculating cell exhibits a slower decay in the correlation function, indicating reduced spin diffusion noise. It is also important to note that the cylindrical cell calculation did not account for beam spot overlap. Since the recirculating design features significantly less overlap, the previously calculated spin correlation coefficient $C_d$ for the cylindrical cell is likely an overestimate.

\section{Conclusion}


In this work, we present a novel recirculating multipass alkali cell designed to overcome spin-noise limitations arising from atomic diffusion in conventional cylindrical and Herriott multipass geometries. We develop a general analytical framework for spin-noise power spectra in multipass cells that explicitly incorporates beam astigmatism and spatial intensity distribution, features not captured in previous models. Using an ABCD matrix approach, we determine beam profiles and extend the diffusion correlation function to multipass configurations, and validate the model against established results for both cylindrical and single-pass geometries. Compared to cylindrical cells, which exhibit non-uniform Lissajous beam patterns and reduced effective filling factors, the recirculating design achieves more uniform spatial coverage and a larger active volume, thereby increasing the number of effectively probed atoms and reducing quantum noise through favorable scaling of $\delta B_{spn}$ and $\delta B_{psn}$ with $(n_v V)^{-1/2}$. We show that spin noise decreases with increasing mirror focal length and that recirculating geometries can outperform conventional cylindrical designs. In addition, selectively suppressing tightly focused regions provides a further route for mitigating diffusion-induced noise. Overall, the proposed recirculating multipass cell offers a practical platform for enhancing sensitivity in precision quantum sensing applications, including atomic magnetometry and optical quantum memories \cite{guo2026highly}.

\section*{CRediT authorship contribution statement}
Qian Ling Kee: Writing – original draft, Investigation, Methodology, Formal analysis.
Lingyi Zhao: Methodology, Formal analysis.
Ruvi Lecamwasam: Formal analysis.
Biveen Shajilal: Methodology, Formal analysis.
Xinan Liang: Software, Validation.
Joel K Jose: Investigation.
Yao Chen: Methodology.
Ping Koy Lam: Resources, Project administration.
Tao Wang: Conceptualization, Supervision, Funding acquisition, Resources, Project administration, Writing – review $\&$ editing, Writing – original draft.

\section*{Data availability statement}
All data supporting the findings of this study are available within the article. Additional information can be provided upon request.

The code used to find the beam positions and width of spots in the recirculating cell is available at GitLab [Online]. More info at: (Recirculating Cell Simulation) \url{https://gitlab.developers.cam.ac.uk/qlk21/recirculating-cell} \cite{Zhao2025}. If you use this code in your work, please cite this paper.

\section*{Acknowledgement}
The authors acknowledge support from A*STAR Career Development Fund (222D800028), Italy-Singapore Science and Technology Collaboration Grant (R23I0IR042), Delta-Q (C230917004, Quantum Sensing), Competitive Research Program (NRF-CRP30-2023-0002),  Q.InC Strategic Research and Translational Thrust and A*STAR National Science Scholarship (PhD). We are grateful for discussions with Ms. Chenyue Gu, Ms. Angela Baiju and Prof. Dong Sheng.

\section*{Declaration of Competing Interest}
The authors declare that they have no known competing financial interests or personal relationships that could have appeared to influence the work reported in this paper.

\setcounter{figure}{0}
\setcounter{table}{0}
\setcounter{equation}{0}
\renewcommand{\thefigure}{A\arabic{figure}}   
\renewcommand{\thetable}{A\arabic{table}}     
\renewcommand{\theequation}{A\arabic{equation}} 



\section*{References}
\bibliography{apssamp}

\end{document}